\documentclass{pnastwo}
\usepackage{PNASTWOF}
\usepackage{amssymb,amsfonts,amsmath,epsfig}
\usepackage{calc}
\usepackage{url}
\usepackage{graphicx}
\begin{document}
\conflictofinterest{Conflict of interest footnote placeholder}
\title{Common group dynamic drives modern epidemics across social, financial and biological domains}


\author{Zhenyuan Zhao\affil{1}{Physics Department, University of Miami, Florida FL 33126, U.S.A.,}, 
Juan Pablo Calder\'on\affil{2}{Industrial Engineering Department, Universidad de Los Andes, Bogota, Colombia},
Chen Xu\affil{3}{School of Physical Science and Technology, Soochow University, Suzhou 215006, People's Republic of China}, 
Dan Fenn\affil{4}{Oxford Centre for Industrial and Applied Mathematics, Oxford University, Oxford OX1 3LB, U.K.}, 
Didier Sornette\affil{5}{ETH Zurich, D-MTEC, Kreuzplatz 5, 8001, Zurich, Switzerland}, 
Riley Crane\affil{5}{}, 
Pak Ming Hui\affil{6}{Department of Physics, Chinese University of Hong Kong, Shatin, Hong Kong}, 
Neil F. Johnson\thanks{To whom correspondence should be addressed. E-mail:
 njohnson@physics.miami.edu}\affil{1}{}
}

\maketitle

\begin{article}

\begin{abstract}
We show that qualitatively different epidemic-like processes from distinct societal
domains (finance, social and commercial blockbusters, epidemiology)
can be quantitatively understood using the same unifying
conceptual framework taking into account the interplay between the
timescales of the grouping and fragmentation of social groups together
with typical epidemic transmission processes. Different domain-specific
empirical infection profiles, featuring multiple resurgences and
abnormal decay times, are reproduced simply by varying the timescales
for group formation and individual transmission. Our model emphasizes
the need to account for the dynamic evolution of multi-connected
networks. Our results reveal a new minimally-invasive dynamical method
for controlling such outbreaks, help fill a gap in existing
epidemiological theory, and offer a new  understanding of complex system
response functions.
\end{abstract}


\noindent  
The world recently witnessed a baffling variety of global outbreak phenomena: the huge fluctuations across the world's financial markets, driven in part by the rapid global spread of rumors\cite{finance}; an unexpected global outbreak of swine flu\cite{flu}, driven in part by rapid social mixing (e.g. within schools\cite{schools,children1}); and even the sudden rise to global fame of an unknown Scottish singer, driven in part by word-of-mouth sharing\cite{boyle,sornette1,riley}. 
To understand these phenomena, consider the following.
The number and identity of the people with whom we are each in instantaneous electronic or physical contact -- and with whom we can exchange information, rumors or viruses -- are characterized by a strong persistence 
related to our social network of acquaintances, coexisting with large intermittent fluctuations arising from random interactions with various social groups. We 
are interested in how the latter may give rise to novel types of dynamics, which are unexplained by standard
epidemic models.
Think for instance of airborne travel in which people remain confined for hours with strangers, unknowingly exchanging 
respiratory pathogens. On the blogosphere and on the Web, ephemeral groups form around topics or content and exchange information, opinions and social contacts before flickering out of existence. 
The transient transnational nature of online discussion groups and chat-rooms, as frequented by financial traders or YouTube users\cite{finance,sornette1,riley} provides a vivid illustration.  

Before dwelling more on real-world examples and illustrating how they might reveal these
novel dynamical properties, we first describe our model which combines group and individual dynamics.
Current epidemiological models have been hugely successful in describing various biological diseases, and in incorporating many realistic details (e.g. spatial topology, differential susceptibility) \cite{Keeling,May,Koopman,cvespignani,schwartz,blasius,dodds}. 
Here, we study the poorly understood \cite{Koopman} dynamical regime shown in Fig. 1, where the group-level dynamics and individual-level transmission processes (modeled by SIR dynamics) evolve on similar timescales -- and hence the number and identity of a given individual's contacts can change abruptly over time.  
In our model (Fig. 1b), individual connectivities may change significantly on the same timescale as that of the SIR process, thereby mimicking those individuals such as participating in YouTube viewing, financial systems, and schools, who may exhibit sometimes rapid moves among peer groups either online or in real space, while simultaneously picking up and spreading rumors or pathogens. 

The grouping process at a given timestep involves coalescence and fragmentation events.
In the simplest implementation, the rate of coalescence of two groups
of size $n_1$ and  $n_2$ respectively  is proportional to the combinatorial number of pairwise encounters
between individuals, one from each group, i.e., the rate of coalescence is equal to $\nu_{coal} \cdot n_1 \times n_2$, where  $\nu_{coal}\leq1$ quantifies the rate of group coalescence per pairs of individuals. Similarly, a given group of 
$n$ individuals may fragment with a total rate equal to $\nu_{\rm frag} \cdot n$, where $\nu_{frag}+\nu_{coal} \leq1$, reflecting the increasing fragility of large groups (standard size effect). 

A discrete illustration over six time steps is shown in Fig. 1b and 1c.
Fig. 1d contrasts the short-time group structure between individuals with the long-term
linkage between them: as time increases without bound, by ergodicity, all individuals will have eventually been 
part of some common group. While the latter long-time network structure is the one usually emphasized
in models of epidemic processes on complex networks,
the short-time limited linkage is essential to understand the competition between individual isolation
(which tends to stop an epidemic) and group coalescence which amplifies its spreading. 
Our model provides a simple framework to quantify the interaction of these group dynamical processes
in combination with the SIR dynamics (with $p$ and $q$ as the infectivity and recovery parameters).

We chose this specific fission-fusion process for several reasons: (1) It embodies the rare but dramatic 
changes of contact networks that can occur, as mentioned in the introduction.
(2)  It produces a distribution of group sizes which is power-law with exponent 5/2 when time-averaged, as shown
in a related model applied to financial markets \cite{ez}, in agreement with many empirical distributions across the natural and social sciences\cite{neil}. (3) The power-law exponent 5/2 is exactly that inferred for group sizes of financial traders, as well as terrorists and insurgent groups, based on an analysis of volume of trades and casualty figures respectively\cite{neil,us}. (4) The model is structurally robust in that the group dynamic rules can be generalized to different positive power exponents $\alpha \neq 1, \beta \neq 1, \gamma \neq 1$, with coalescence and fragmentation rates given by
$\nu_{coal} \cdot n_1^{\alpha} \times n_2^{\beta}$ and  $\nu_{\rm frag} \cdot n^{\gamma}$, respectively,
without losing the main qualitative features of the dynamics of the number $I(t)$ of infected individuals.

In the numerical implementation of the model, we run the above coalescence-fragmentation dynamics until
the time-averaged distribution of group sizes has become stationary. Then, at some
instant taken at the origin of time $t=0$, one group is selected, and an arbitrary individual in this group becomes infected, and hence the infection profile unfolds according to the SIR process within each group, with all the groups
undergoing at the same time the coalescence-fragmentation dynamics according to the two rates $\nu_{coal}$
and $\nu_{\rm frag}$. Our model describes genuine within-group dynamics 
coexisting with between-group dynamics (our model is 
fundamentally different from a homogeneous-mixing metapopulation type model).

A typical simulation is shown in Figure 2, and is compared with
the popular approach that models spreading on static networks: (i) an instantaneous network
($T=1$, purple curve) and (ii) a global network formed by time-aggregating instantaneous contacts
over long times ($T\rightarrow \infty$, green curve). SIR spreading dynamics on fixed networks
obtained at different intermediate $T$ gives curves that lie in the shaded area of Figure 2. 
Our model can generate not only this type of dynamics, but also qualitatively new regimes that arise from adjusting the coalescence-fragmentation rates:
the large fluctuations, resurgences, and abnormally long decay time which are observed in our model 
(and illustrated in Figure 2 (blue curve)) are generated by self-amplification and suppression processes 
due to the fission-fusion group dynamics at all group-size scales.

How does this compare with real-life epidemic-like dynamics?
The top two rows of Figure 3 show two typical example datasets that we have collected in each of the three domains mentioned above. (See below, and supporting online material \url{http://www.er.ethz.ch/publications/complex_systems/internet/group_dynamics_som.pdf}, for experimental details). These three
examples have been chosen to illustrate bursts of activity followed by slow intermittent relaxations, which can not be 
accounted for by standard SIR dynamics on fixed network topologies. The third row shows a realization
of our model with sets of parameters chosen so that the main properties of the two first rows are 
qualitatively reproduced. These comparisons illustrate the power of our model to account for 
quantitatively different regimes in distinct domains within the same unifying conceptual framework
taking into account the interplay between the
timescales of the grouping and transmission process.

The left panel of Fig. 3 shows downloads for two similar YouTube clips\cite{sornette1,riley}. Such downloads are typically driven by YouTube users absorbing and spreading opinions as they share information in their social groups\cite{riley}. 
The two downloads appealed to similar age-groups, and were measured close together in time, implying that a similar pool of users accessed them both, in line with our model's assumptions. The fission-fusion group dynamics 
with SIR epidemic spreading accounts well for the long memory and aftershock-like decay.
The middle panel shows foreign exchange movements as a result of a specific rumor spreading among traders concerning revaluation of the Chinese Yuan currency. This same rumor circulated twice in the space of a few months. The fact that the currency pairs follow a similar dynamical pattern in each case, suggests that the same 
underlying group dynamics developed, in line with our model.  Note that this
financial epidemic is characterized by 
the largest coalescence rate $\nu_{\rm coal}$ and much larger infectivity parameter $p$
among the three examples, reflecting
the efficiency of the information cascade among currency traders.
The right panel shows incidences of a cold among 1st grade students in two schools in Bogota, Colombia. The schools' location guarantees that seasonal temperature variations are minimal, and the student population of each approximates to a closed system due to local issues of security and social segregation.  We argue that the immune systems of the children have been subjected to a soup of
microbes coming and going, so that the successive bursts are part of the same dynamics, especially
given the unchanging climatic conditions of this part of Columbia.
Within our approach, the school cold dynamics
are found to be best described by the lowest fragmentation rate $\nu_{\rm frag}$ and highest recovery parameter $q$, mirroring the more rigid structure of inter-children contacts and the crucial role of multiple recuperations.

A full analytic description of $I(t)$ represents a fascinating open challenge.  However some features can be captured by suitably generalizing existing epidemiological machinery. A key quantity is the probability $\eta_{\rm SI}(t)$ that a particular link instantaneously exists {\em and} that it connects a susceptible and an infected. Thus, out of a potential totality of $N(N-1)/2$ links among $N$ individuals, only $\eta_{\rm SI}(t) \cdot N(N-1)/2$ are typically present. This 
provides an accurate equation for the number of susceptibles $S(t)$ in a given epidemic sequence: ${\dot S}(t)=-p.\eta_{\rm SI}(t).N(N-1)/2$, with $p$ is the infectivity parameter quantifying the 
probability that a particular $S$ is infected by a particular $I$ to which it is linked. 
We rewrite this in the conventional mass-action form ${\dot S}(t)=-p P_{\rm SI}(t) S(t) I(t)$, where $P_{\rm SI}(t)$ is equivalent to $\eta_{\rm SI}(t) N(N-1)/[ 2 S(t) I(t)]$ and hence now incorporates the complex dynamics which are so hard to capture analytically. We now approximate $P_{\rm SI}(t)$ by a constant term $P$, which is the time-averaged probability that any two arbitrarily chosen nodes belong to the
same cluster independent of $SI$-infection status. This approximation throws out the dynamical details of $I(t)$, but can provide useful insights, as shown below on the non-spreading to spreading transition, provided that several coalescence-fragmentation processes occur over the timescale of the entire outbreak. We obtain $P$ by solving a time-dependent master equation, yielding $P=\nu_{coal}/(N\nu_{frag})$ if the coalescence and fragmentation probabilities are of the same order and $N\gg 1$. This result can be derived independently using generating function techniques on the coupled equations for groups of all sizes. 

We now use this result to address a highly topical, yet previously unaddressed, question:  {\em Will there be epidemic spreading in a population in which it is publically known that $N_0$ persons have been infected with a given pathogen or rumor, but where the precise identity of infected persons cannot be disclosed?} At $t=0$, $N_0 \ll N$ individuals of
the instantaneously largest group are infected and news of an infection is announced without disclosing the infected's identities. The population reacts by adjusting its group dynamics, i.e. it adopts a new $\nu_{\rm coal}$ and $\nu_{\rm frag}$.  The number of susceptibles in the long-time limit $S(\infty)$ with $N\gg 1$ is then given by the solution $\overline{z}$ to the following generalized form, 
$z = {\rm exp}{[- \kappa (1 - z)]}$ where $z \equiv S(\infty)/N$ and $\kappa \equiv p\nu_{\rm coal} /q\nu_{\rm frag}$.
For $\kappa \leq 1$, the only solution is  $\overline{z} = 1$, corresponding to a vanishingly small
fraction of infected individuals (i.e.  total number of infected $R(\infty)$ does not exceed $N_0 \ll N$). This solution
bifurcates at $\kappa =1$ into the following stable solution
$\overline{z}= -(1/\kappa) \cdot W\left(-\kappa \cdot e^{-\kappa}\right)$  valid for $\kappa >1$, where 
$W(z)$ is the Lambert function. For $\kappa>1.5$,  $\overline{z}$ is very well-approximated
by $\overline{z} \approx e^{-\kappa}/\kappa$. This shows a rather abrupt transition from
non-spreading epidemics for $\kappa < 1$ to global infection of a finite fraction of the population
for $\kappa >1$.  The form of the epidemic control parameter 
$\kappa \equiv p\nu_{\rm coal} /q\nu_{\rm frag}$ exemplifies that infectivity and coalescence
play together against recovery and fragmentation in controlling the propagation of the epidemics:
Infectivity and coalescence promote the infection propagation, while recovery and fragmentation
hinders its spread. Our model provides a new window of opportunity for controlling epidemics,
by providing a quantitative framework to better target
actions at the group level through its fission-fusion dynamics, thus improving the distinction
with actions at the
the more intrinsic level of the host-virus characteristics and dynamics $(p,q)$.

Figure 4 shows the resulting phase diagram as a function of the new $\nu_{\rm coal}$ and $\nu_{\rm frag}$. Not only is our theory for the spreading threshold (dashed black line in Fig. 4) in excellent agreement with the numerical results (white solid line), its simple analytic form suggests a novel yet generic epidemic control scheme based on manipulation of the group coalescence and fragmentation timescales (i.e. $\nu^{-1}_{\rm coal}$ and $\nu^{-1}_{\rm frag}$). An imminent epidemic can be  {\em suppressed} (i.e. $R(\infty)<N_0$) by increasing the timescale for group coalescence  with respect to the timescale for group fragmentation (i.e. decrease $\nu_{\rm coal}$ with respect to $\nu_{\rm frag}$), but it will get {\em amplified} if we decrease the coalescence timescale with respect to the fragmentation timescale (i.e. increase $\nu_{\rm coal}$ with respect to $\nu_{\rm frag}$). Not only would such modest intervention allow the overall system to continue functioning, it does {\em not} require knowledge of the infected's identities. There is also no assumption that the $N_0$ members of the group which carries the initial infected case at $t=0$, remain in that group. In the school setting, schedules could be adjusted to slow down or speed up classroom use and recess, without the need for disruptive school closures\cite{children1} or the need to test, label or isolate infected children. Similar control can be achieved in the online chatrooms frequented by financial traders, by basing the joining and leaving rules on present occupancy. In viral marketing, the attractiveness of the message or product quantified by the infectivity $p$ can be completely
subjugated by suitable management of the group dynamics ($\nu_{\rm coal}$ versus $\nu_{\rm frag}$), as firms using e-commerce and e-advertisement are now
realizing. These findings are potentially applicable to many other scenarios, given that many real-world activity/infection curves resemble those in Figure 2.  Our framework offers a theoretical underpinning to formulate new solutions as well as getting new sources
of data probing the inner working of the interplay between within-group 
and between-group dynamics. 

\vskip 0.3cm

\noindent {\bf Data}  Figure 3, left panel: `Gettin' Enough' from 
\url{http://www.youtube.com/watch?v=AiXxMrkeklg}.
Video uploaded to site: Wed, 08 Nov 2006 13:33:12 GMT. First record of download:   Thu, 09 Nov 2006 17:21:35 GMT, view count: 5708.
Last record:    Thu, 24 May 2007 21:34:30 GMT,  view count: 257759.
Length of video: 225 seconds.
Middle-left panel, downloads of music video `Borat' from  
\url{http://www.youtube.com/watch?v=b1xXERFt_Zg}.
Video uploaded to site:  Fri, 03 Nov 2006 11:04:15 GMT. First record of download:  Tue, 07 Nov 2006 10:35:57 GMT, view count: 20745.
Last record:   Thu, 24 May 2007 22:29:25 GMT,   view count: 254918.
Length of video: 154 seconds. These two music downloads are similar in terms of appeal, age-group, total number of downloads, and lack of any public/global announcement, news or advertisement, hence consistent with spreading through contagion.
Bottom-left panel, model output with $\nu_{\rm frag}=0.05$, $\nu_{\rm coal}=0.81$, $p=0.001$, $q=0.001$. Financial exchange rates for currency (centre panels):
Top-centre panel, CNY (Chinese currency) revaluation rumor, detected from trader chat-rooms by HSBC bank (courtesy of S. Williams). Absolute returns on the timescale of 1-minute intervals, JPY (Japanese currency) exchange rates from 08:22 to 08:53 GMT, on 11 May 2005.
Middle-centre panel: CNY actual revaluation.
Absolute returns on the timescale of 1-minute intervals, JPY (Japanese currency) exchange rates from 11:03 to 11:34 GMT, on 21
July 2005.
Since the CNY was not one of the directly traded currencies, its effects on the JPY-X rates (where X is another currency) are indirect in both cases, suggesting influence through contagion of the rumor/information. There was no public announcement or global news to trigger this activity, which also supports spreading through contagion.
Bottom-centre panel: model output with  $\nu_{\rm frag}=0.05$, $\nu_{\rm coal}=0.95$, $p=0.009$, $q=0.002$. Since data are on 1-minute scale, but prices can change on 1-second scale, we show an averaged output by providing value at regular equispaced intervals, mimicking 1-minute.
School colds data (right panels): 
Fraction of 1st grade children with colds
in Marymount School and CNG School, Bogota, Colombia. Bottom-right panel: model output with  $\nu_{\rm frag}=0.001$, $\nu_{\rm coal}=0.5$, $p=0.001$, $q=0.004$. The model output is smoothed, to mimic fact that data is recorded on the 1-week time interval.

\renewcommand\refname{References and Notes}

\end{article}

\begin{figure*}[hb]
\centering
\includegraphics[width=0.8\textwidth]{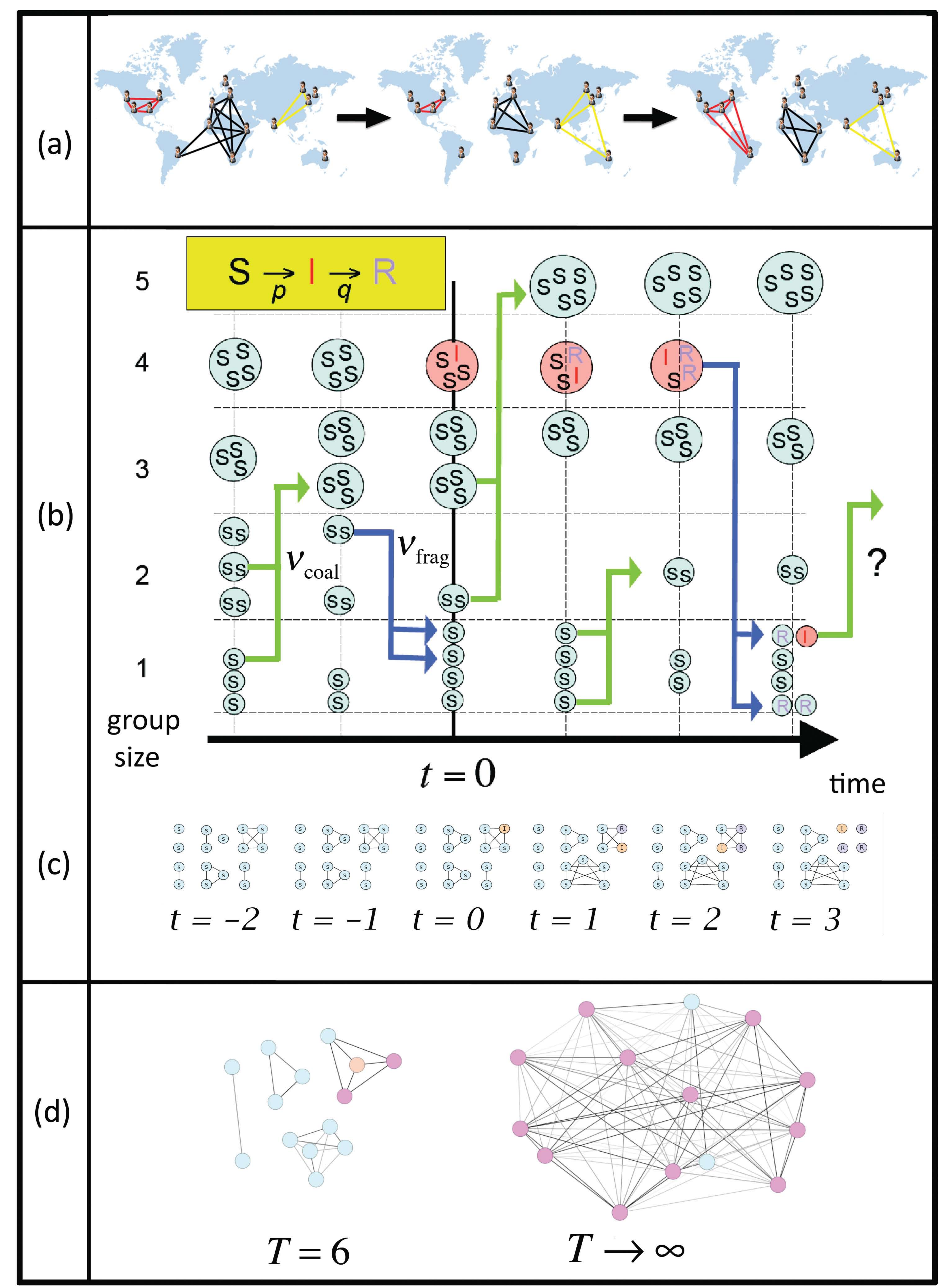}
\caption{Our dynamical group contagion model. 
a: Schematic of dynamical grouping of traders or YouTube users on the Internet
b: Spreading in the presence of dynamical grouping via coalescence and fragmentation.  Vertical axis shows number of groups of a given size at time $t$. c: Instantaneous network from Fig. 1b at each timestep. 
d: Weighted network obtained by aggregating links over time-window $T$.}
\end{figure*}

\begin{figure*}[hb]
  \centering
\includegraphics[width=1\textwidth]{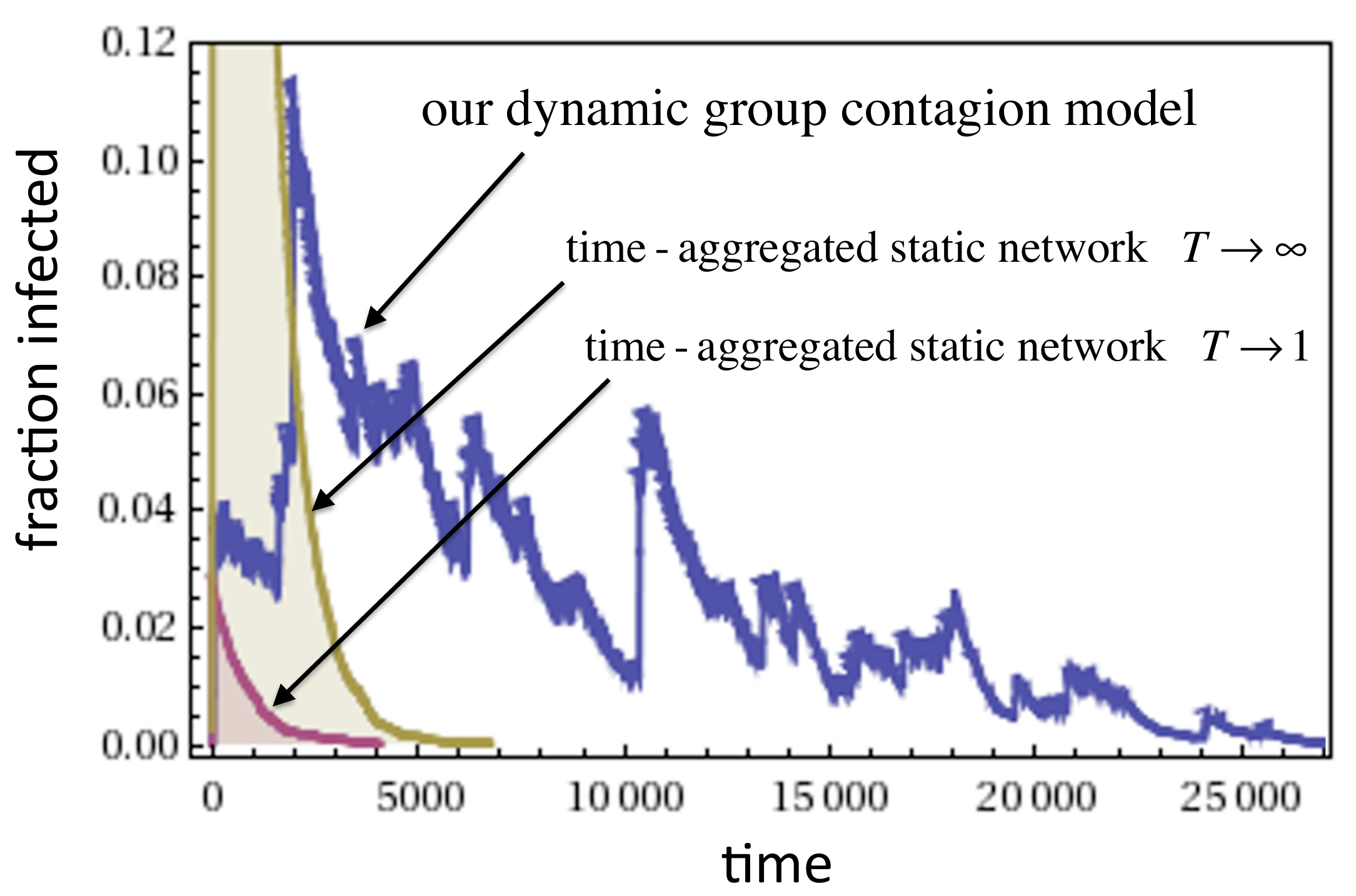}
\caption{Theoretical profile $I(t)$. Blue curve shows our dynamical group contagion model, with $\nu_{\rm frag}=0.05$, $\nu_{\rm coal}=0.95$, $p=0.001$ and $q=0.001$. Using same $p$ and $q$ values, purple curve corresponds to stochastic SIR model on a static network with $T\rightarrow 1$ (i.e. the $t=0$ network in Fig. 1). Green curve corresponds to stochastic SIR on a $T\rightarrow \infty$ network.}
\end{figure*}

\begin{figure*}[hb]
  \centering
	\includegraphics[width=1.0\textwidth,angle=270]{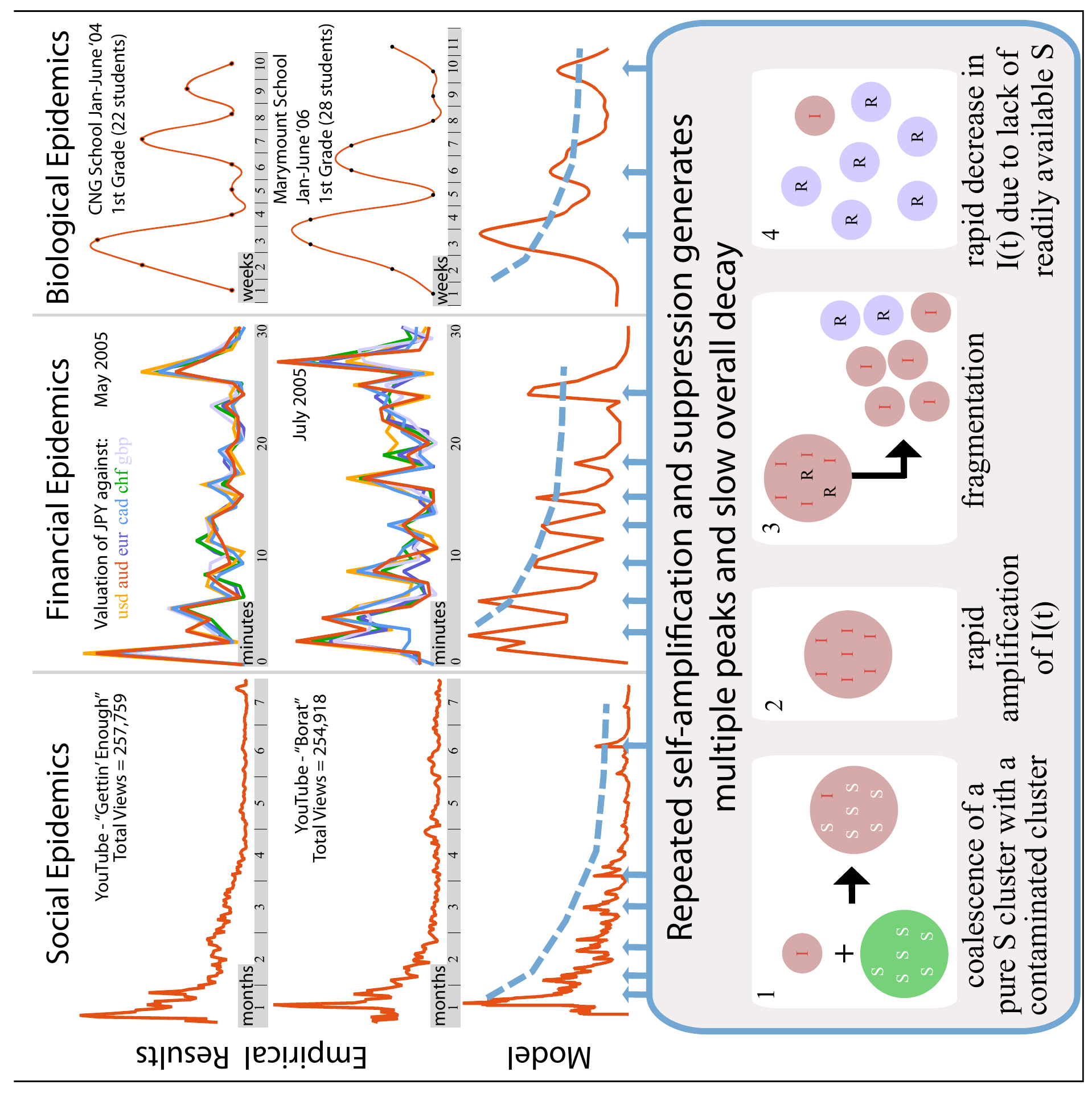}
\caption{Modern epidemics. Top two rows: Empirical activity profile $I(t)$ in three distinct real-world systems. Third row: Results from dynamical grouping model. Left: YouTube download activity. Middle: Currency trading activity (i.e. absolute value of price-change, hence the excess demand to buy or sell at each timestep). Right: Fraction of children with colds within a school. Lower panel: Simple example of the repeated self-amplification and suppression processes which spontaneously arise within our the model. When replicated at all scales of group size, these processes generate a unified quantitative description of the empirical $I(t)$ profiles.}
\end{figure*}

\begin{figure*}[hb]
  \centering
	\includegraphics[width=1\textwidth]{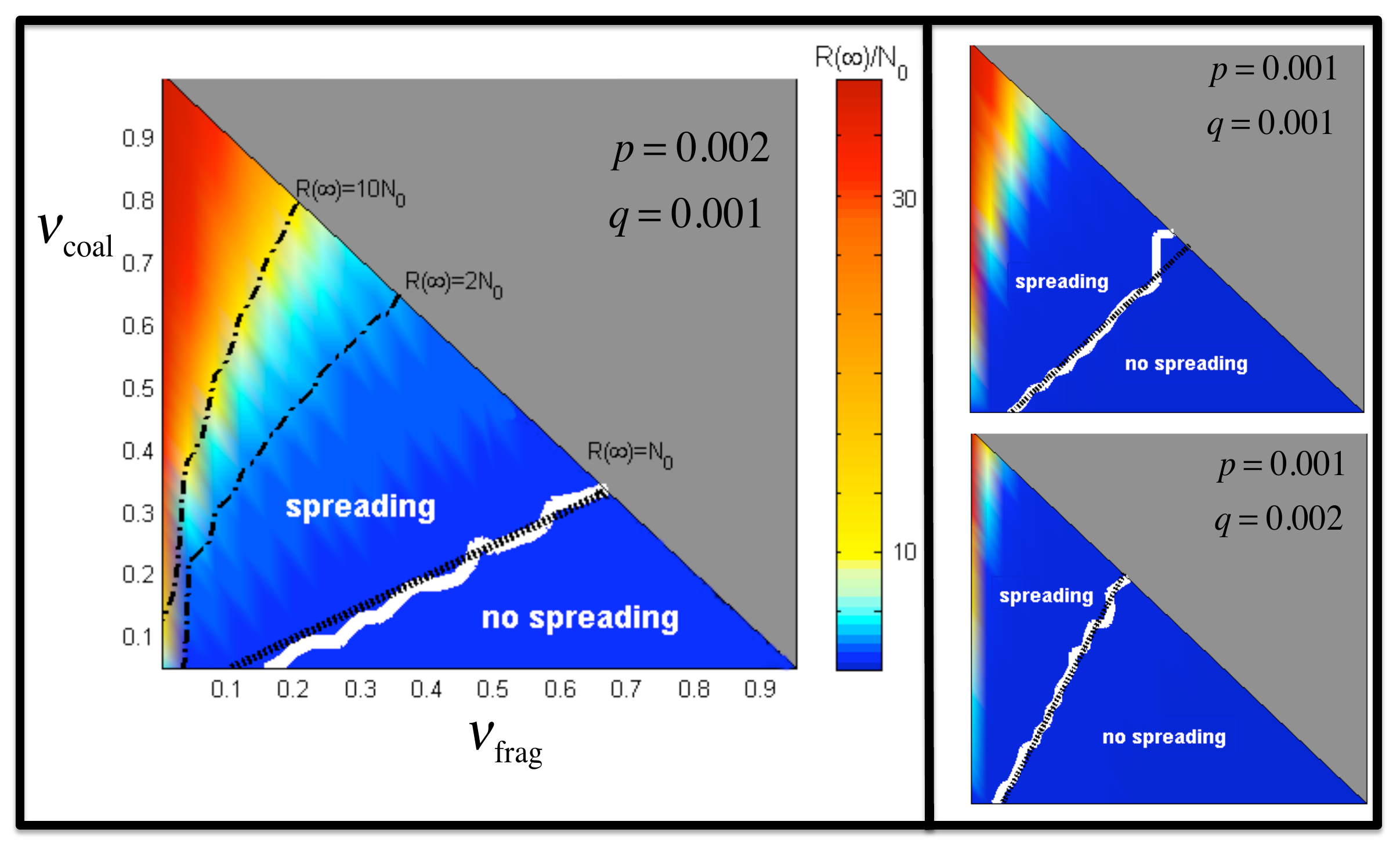}
\caption{Consequences of human reaction to news of an outbreak.
Phase diagrams show theoretically obtained transition (i.e. $\frac{p\nu_{\rm coal}}{q\nu_{\rm frag}} = 1$, black dashed line) and the numerical result (white line) separating regimes of spreading (i.e. overall number of infecteds exceeds initial group size, hence $R(\infty)>N_0$) and no-spreading (i.e. $R(\infty)<  N_0$). Population reacts to news of the initial infection at $t=0$ by changing its dynamical grouping from $\nu_{\rm frag}=0.001$ and $\nu_{\rm coal}=0.99$, to the new values shown on the axes. Colours show the population (in units of $N_0$) who become infected, and hence recovered, over the lifetime of the outbreak.  Shaded grey region is unphysical since $\nu_{\rm frag}+\nu_{\rm coal}>1$.}
\end{figure*}

\end{document}